\documentclass{aa}  
\usepackage{graphicx,color}
\usepackage{txfonts}
\usepackage{amsmath}
\usepackage{xcolor}
\usepackage{natbib,twoopt}
\usepackage[hyphenbreaks]{breakurl}
\usepackage[breaklinks]{hyperref}  
\DeclareMathAlphabet\mathzapf       {T1}{pzc} {mb} {it}
\bibpunct{(}{)}{;}{a}{}{,}             
\definecolor{cobalt}{rgb}{0.06, 0.2, 0.65}
\usepackage{hyperref}
\hypersetup{
  colorlinks,
  citecolor=cobalt,
  linkcolor=cobalt,
  urlcolor=cobalt,
}
\makeatletter
  \newcommandtwoopt{\citeads}[3][][]{\href{http://adsabs.harvard.edu/abs/#3}%
    {\def\hyper@linkstart##1##2{}%
     \let\hyper@linkend\@empty\citealp[#1][#2]{#3}}}
  \newcommandtwoopt{\citepads}[3][][]{\href{http://adsabs.harvard.edu/abs/#3}%
    {\def\hyper@linkstart##1##2{}%
     \let\hyper@linkend\@empty\citep[#1][#2]{#3}}}
  \newcommandtwoopt{\citetads}[3][][]{\href{http://adsabs.harvard.edu/abs/#3}%
    {\def\hyper@linkstart##1##2{}%
     \let\hyper@linkend\@empty\citet[#1][#2]{#3}}}
  \newcommandtwoopt{\citeyearads}[3][][]%
    {\href{http://adsabs.harvard.edu/abs/#3}
    {\def\hyper@linkstart##1##2{}%
     \let\hyper@linkend\@empty\citeyear[#1][#2]{#3}}}
\makeatother
\newcommand{\be}{\begin{equation}}
\newcommand{\en}{\end{equation}}

\def\ltsima{$\; \buildrel < \over \sim \;$}
\def\lsim{\lower.5ex\hbox{\ltsima}}
\def\gtsima{$\; \buildrel $\geq$ \over \sim \;$}
\def\gsim{\lower.5ex\hbox{\gtsima}}

\begin{document} 

   \title{Fast X-ray/IR observations of the black hole transient Swift~J1753.5--0127: from an IR lead to a very long jet lag}
\titlerunning{Fast IR photometry of Swift~J1753.5--0127}
\authorrunning{Ulgiati et al.}

   \author{A. Ulgiati
          \inst{1,2}, F. M. Vincentelli\inst{3,4}, P. Casella\inst{5}, A. Veledina\inst{6, 7}, T. J. Maccarone\inst{8}, D. M. Russell\inst{9}, P. Uttley\inst{10}, F. Ambrosino\inst{5}, M. C. Baglio\inst{11}, M. Imbrogno\inst{12,5}, A. Melandri\inst{5}, S. E. Motta\inst{11,13}, K. O'Brien\inst{14}, A. Sanna\inst{15}, T. Shahbaz\inst{3,4}, D. Altamirano\inst{16}, R. P. Fender\inst{13}, D. Maitra\inst{17}, J. Malzac\inst{18}
          }

   \institute{INAF - IASF Palermo, via Ugo La Malfa, 153, I-90146, Palermo, Italy \\
              \email{alberto.ulgiati@inaf.it}
         \and
             Universit\`{a} degli Studi di Palermo, Dipartimento di Fisica e Chimica, via Archirafi 36, I-90123 Palermo, Italy
         \and 
             Instituto de Astrof\'{i}sica de Canarias, E-38205 La Laguna, Tenerife, Spain
         \and
         Departamento de Astrof\'{ı}sica, Universidad de La Laguna, E-38206 La Laguna, Tenerife, Spain
         \and
            INAF – Osservatorio Astronomico di Roma, via Frascati 33, I-00078 Monteporzio Catone, Italy
         \and
             Department of Physics and Astronomy, FI-20014 University of Turku, Finland
         \and    
             Nordita, Stockholm University and KTH Royal Institute of Technology, Hannes Alfv\'{e}ns v\"{a}g 12, SE-10691 Stockholm, Sweden
         \and
             Department of Physics \& Astronomy, Texas Tech University, Box 41051, Lubbock, TX, USA, 79409-1051
          \and
             Center for Astrophysics and Space Science (CASS), New York University Abu Dhabi, P.O. Box 129188, Abu Dhabi, UAE
          \and
             Anton Pannekoek Institute, University of Amsterdam, Science Park 904, 1098 XH Amsterdam, The Netherlands
          \and
             INAF, Osservatorio Astronomico di Brera, via E. Bianchi 46, I-23807 Merate (LC), Italy
          \and
             Dipartimento di Fisica, Università degli Studi di Roma “Tor Vergata”, via della Ricerca Scientifica 1, I-00133 Rome, Italy
          \and
             University of Oxford, Department of Physics, Astrophysics, Denys Wilkinson Building, Keble Road, OX1 3RH, Oxford, United Kingdom
          \and 
             Centre for Advanced Instrumentation, Department of Physics, Durham University, Durham, UK
          \and
             Dipartimento di Fisica, Università degli Studi di Cagliari, SP Monserrato-Sestu km 0.7, 09042 Monserrato, Italy
          \and
             School of Physics and Astronomy, University of Southampton, Southampton, Hampshire SO17 1BJ, UK
          \and
             Department of Physics and Astronomy, Wheaton College, Norton, MA 02766, USA
          \and
             IRAP, Universit\'{e} de Toulouse, CNRS, UPS, CNES, Toulouse, France
             }

   \date{Received ...; accepted ...}
 
\abstract{
We report on two epochs of simultaneous near-infrared (IR) and X-ray observations with a sub-second time resolution of the low mass X-ray binary black hole candidate Swift J1753.5--0127 during its long 2005--2016 outburst. Data were collected strictly simultaneously with VLT/ISAAC (K$_{S}$ band, 2.2 $\mu m$) and RXTE (2-15 keV) or \textit{XMM-Newton} (0.7-10 keV).
A clear correlation between the X-ray and the IR variable emission is found during both epochs but with very different properties. In the first epoch, the near-IR variability leads the X-ray by $ \sim 130 \, ms$. This is the opposite of what is usually observed in similar systems. The correlation is more complex in the second epoch, with both anti-correlation and correlations at negative and positive lags. Frequency-resolved Fourier analysis allows us to identify two main components in the complex structure of the phase lags: the first component, characterised by a few seconds near-IR lag at low frequencies, is consistent with a combination of disc reprocessing and a magnetised hot flow;  the second component is identified at high frequencies by a near-IR lag of $\approx$0.7 s. Given the similarities of this second component with the well-known constant optical/near-IR jet lag observed in other black hole transients, we tentatively interpret this feature as a signature of a longer-than-usual jet lag. We discuss the possible implications of measuring such a long jet lag in a radio-quiet black hole transient.
}


   \maketitle
%
%-------------------------------------------------------------------

\section{Introduction}

Black hole transients (BHTs) are low-mass X-ray binaries (LMXBs) that show long periods of quiescence interrupted by shorter periods of activity (weeks to years) called outbursts \citep{2006ARA&A..44...49R}. During such events, these systems show strong and variable emission over a large part of the electromagnetic spectrum, from the radio to the hard X-rays. Three main components have been identified to contribute to this multi-wavelength emission. Thermal emission from an irradiated accretion disc is believed to be responsible for the emission from soft X-rays to the optical-infrared (O-IR) band. \citep{1973A&A....24..337S}. Hard X-ray photons are associated with inverse Compton scattering by a population of energetic electrons, often referred to as a "corona" \citep{1997ApJ...489..865E,1997MNRAS.292L..21P}. Arguments involving the energetics of the corona indicate it must be located in the innermost regions of the accretion flow, although its actual geometry is still a matter of debate \citep{2007A&ARv..15....1Done,2018A&A...614A..79Poutanen,2021SSRv..217...65Bambi}. Some models assume the corona is magnetised, which causes further emission at lower energy, e.g. in the optical or even infrared band, via synchrotron emission \citep{2000MNRAS.318L..15M}. Finally, steady, compact jets --  collimated streams of matter ejected in the direction orthogonal to the accretion plane at nearly relativistic speeds \citep{1979ApJ...232...34B,2001MNRAS.322...31Fender} - are also observed with a typical synchrotron flat spectrum which extends from radio to O-IR wavelength \citep{1988ApJ...328..600H,2002ApJ...573L..35C}.

During their outbursts, BHTs show two main spectral states: an X-ray \textit{hard state}, where the highly variable emission is dominated by the high-energy X-ray photons emitted by the corona, and an X-ray \textit{soft state}, where the stable low-energy X-ray thermal emission from the disc dominates the X-ray spectrum \citep{2011BASI...39..409B}. A jet is always observed when a source is in its hard state, while no compact radio source has ever been detected in the soft state \citep{2020MNRAS.498L..40Maccarone}, suggesting that the jet is quenched, unless it has changed drastically its emissivity properties (\citealt{2009ApJ...703L..63C};\citealt{2017MNRAS.466.4272D}; but see \citealt{2018A&A...612A..27Koljonen})

Most BHTs show a similar evolution:  they begin their outburst in the hard state \citep{2005A&A...440..207B}, and keep roughly the same hardness as the luminosity increases. Then they undergo a transition from hard to soft state. During this transition, the source goes through a so-called hard-intermediate state, in which most of the characteristic timescales of the X-ray variability decrease, and the emission from the steady compact jet is observed to quench, and then through a short-lived soft-intermediate state, to which discrete, powerful ejections are often associated, right before entering the soft state \citep[e.g.][]{2009MNRAS.396.1370F}. When in the soft state the source luminosity declines nearly steadily until a transition back to the hard state is observed, during which the emission from the compact jet reappears and increases until, eventually, the source heads back to quiescence at all wavelengths \citep{2013MNRAS.431L.107C_corbel}.
While this pattern is observed regularly in most BHTs, some outbursts do not go through the complete cycle, as they never reach the soft state (or even leave their hard state), before starting their decline towards quiescence \citep[the so-called "failed-transition outbursts", see e.g. ][and references therein]{2004NewA....9..249B,2021MNRAS.507.5507A}.

Large-amplitude variability can be observed at all wavelengths on different timescales, depending on the state. The different emitting components are necessarily inter-connected through the inflowing/outflowing matter itself and through irradiation, thus studying the correlation between the variability at different wavelengths plays a key role in understanding the emission mechanisms, in measuring the physical parameters of the system and investigating the links between the various emitting regions. In particular, the study of the correlation between the multi-wavelength emissions at high time resolution allows us to probe the regions in the immediate vicinity of the compact object \citep{2004PThPS.155...99Z,Gilfanov_2009,2006ARA&A..44...49R,2014SSRv..183...43B,2014SSRv..183...61P}.

The development of high quantum efficiency fast optical-infrared photometers opened the possibility of studying the fast variability from these systems at lower energies and linking it to the behaviour in X-rays. After a handful of pioneering works in the 1980s \citep{1982A&A...109L...1M,1983A&A...119..171M}, the first X-ray optical cross-correlation study of XTE J1118+480 revealed the presence of a complex connection between the two bands \citep{2001Natur.414..180K}. The shape of the cross-correlation function showed an anti-correlation at negative lags (the so-called \textit{precognition dip}) followed by a long response at $\sim 8$~s, which was interpreted with the presence of a common energy reservoir between the (X-ray emitting) corona and the (optical) jet \citep{2004MNRAS.351..253M}. The presence of an anti-correlation between X-ray and O-IR as well as long optical responses ($\sim$few s) have been observed in a handful of other sources, e.g. Swift J1753.5--0127 \citep{2008ApJ...682L..45D_durant,2011MNRAS.410.2329D_durant}, MAXI J1535--571 \citep{2021MNRAS.503..614V_vincentelli}, MAXI~J1820+070 \citep{2021MNRAS.505.3452P_paice}. Alternative models have been proposed to explain this behaviour. One of the most successful is the so-called ``extended hot flow model'', which assumes that the optical arises from synchrotron radiation from the external regions of a magnetised corona, while the X-rays arise from synchrotron self-Compton emission \citep{2011ApJ...737L..17Veledina}.

Another common feature which has been observed in these systems is a narrow, symmetric $\approx$0.1~s lag between the X-ray and the O-IR emission \citep{2010MNRAS.404L..21C,2010MNRAS.407.2166G,2017NatAs...1..859G_gandhi}. Given its properties, it is commonly accepted that such a feature is the result of mass accretion rate fluctuations (emitted in X-ray) injected in the jet and re-emitted as synchrotron radiation, possibly through the formation of shocks \citep{2014MNRAS.443..299M_2014,2018MNRAS.480.2054M_malzac,2018MNRAS.477.4524V_vincentelli,2019ApJ...887L..19V_vincentelli,2019MNRAS.490L..62P_paice,2021MNRAS.504.3862T_tetarenko}.

High-cadence, evenly-sampled data permits Fourier domain (cross-)spectral analysis on these systems, leading to the discovery of O-IR quasi-periodic oscillations  \citep[QPOs,][]{1983A&A...119..171M,2010MNRAS.407.2166G,2016MNRAS.460.3284K,2019ApJ...887L..19V_vincentelli}.  QPOs from LMXBs have been studied in X-rays for decades, but their origin is still debated \citep[see e.g.][and references therein]{2019NewAR..8501524I}. The properties of the most commonly observed QPOs \citep[also known as type C, see][and references therein]{2005ApJ...629..403C} have been found to depend on binary inclination \citep{2015MNRAS.447.2059M}. This supports their interpretation in terms of a precessing inflow. In this scenario, the optical-infrared counterparts of type-C QPOs have been described in terms of synchrotron emission from a jet precessing together with the hot flow, or from the (magnetised) hot flow itself.

\begin{figure*}
\centering 
\includegraphics[width=1\textwidth]{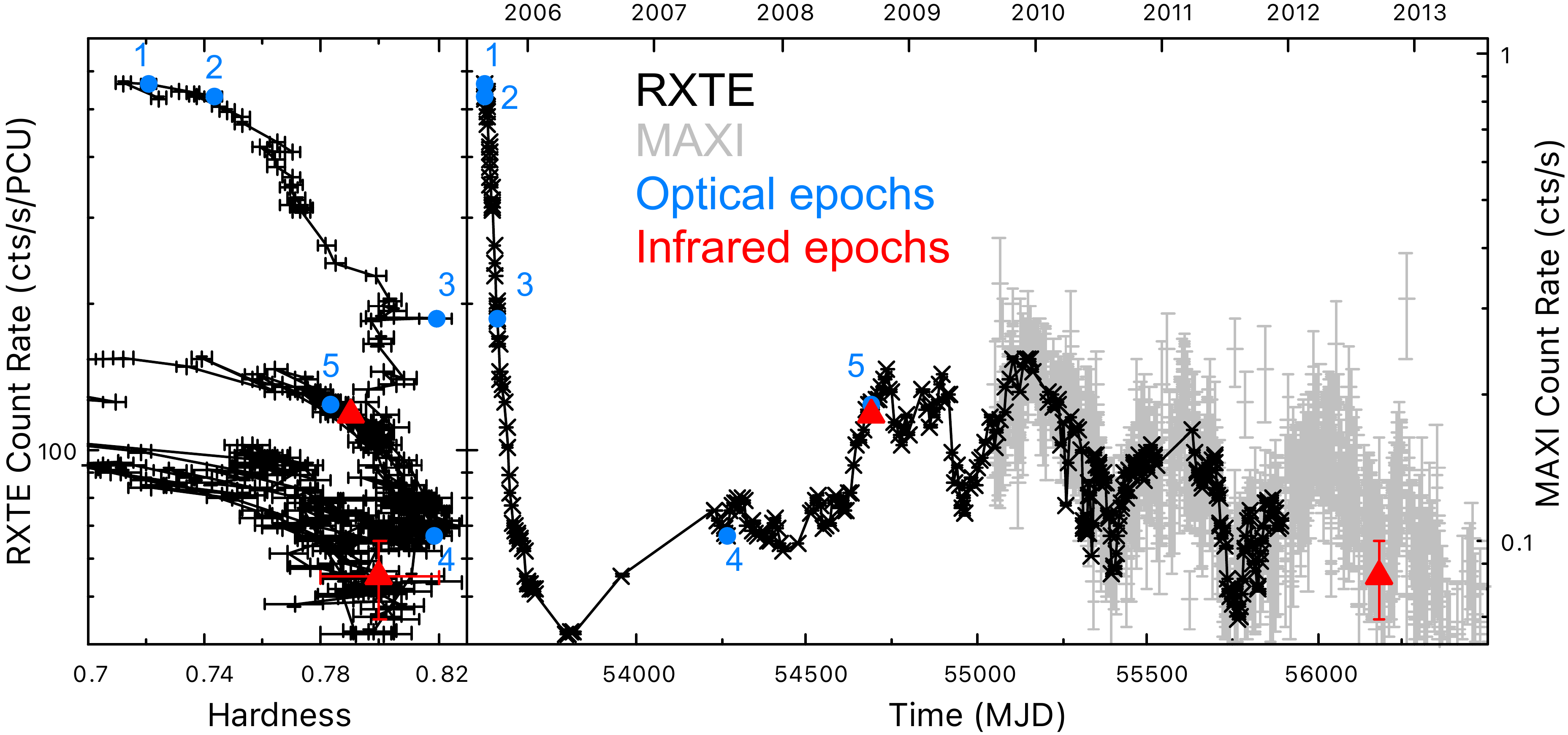}
\caption{Hardness-intensity diagram (HID, left) and light curve (right) of the BHT Swift J1753.5--0127 during its $\sim 10$ year-long outburst that started in 2005. The black points represent the RXTE/PCA data: the count rate is in the energy range $2-15$~keV, while the hardness is the ratio between the rates in the $4-9$~keV and the rates in the $2-4$~keV energy ranges. The grey points represent the MAXI data ($2-15$~keV). The blue points indicate the five epochs of optical fast photometry considered in \citet{2017MNRAS.470...48V}, while the red triangles indicate the two epochs of IR fast photometry reported in this work. The position of the second IR epoch in the HID and its error bars were estimated from the MAXI scaled count rate and from the ratio (not reported here) between the Swift/BAT and the MAXI scaled count rates and should be considered only as indicative of the approximate position of the source in the HID in that epoch, given the low signal-to-noise ratio of both the MAXI and BAT data. Note that the soft excursions of the source in the HID (reaching values as low as 0.3) have not been plotted for clarity of visualization \citep[see][for a full-scale representation of the HID]{2019MNRAS.487.1439B}.} 
\label{fig:light_curve}
\end{figure*}

Most of the interpretative efforts for the observed O-IR/X-ray fast variability of LMXBs have been based so far on single observing epochs, given the scarcity of multi-epoch campaigns. This has limited the possibility of linking together the different observed behaviours in a single interpretative scenario. One of the best exceptions to this is represented by Swift J1753.5--0127. This transient was discovered in June 2005 when it started its first outburst, lasting about 10 years
\citep[][and references therein; Fig. \ref{fig:light_curve}]{2010MNRAS.406.1471S,2013MNRAS.429.1244S,2017ApJ...848...92P,2017ApJ...850...92D,2019MNRAS.487.1439B}. The source remained most of the time in the hard state, with occasional excursions into the hard-intermediate state, and a short-lived reported excursion into the soft state \citep[][Fig. \ref{fig:light_curve}, left panel]{2016MNRAS.458.1636S}. The discovery of a 3.24hr likely super-hump modulation \citep{2008ApJ...681.1458Z} suggests that the system has one of the shortest orbital periods among BHTs \citep[][]{Corral-Santana_2016, Tetarenko_2016}.

%\subsection{The BHT Swift J1753.5-0127}
%It was one of the first black-hole transients to be classified as an "outlier" in the radio/X-ray plane, as its low radio flux places it below the so-called "standard" track (REFs). A very short orbital period was reported, making it one of the most compact black-hole binaries (REF). Several epochs of fast simultaneous optical/X-ray photometry were reported for this source, revealing a complex scenario, with an evolving correlation between the two bands.

Given its long and peculiar outburst, this system has been the target of several O-IR and X-ray observations \citep[][and references therein]{2017MNRAS.470...48V}, which permitted the study of the evolution of the X-ray/optical fast variability. Such evolution was well reproduced by the aforementioned hot-inflow model, suggesting an evolution of the flow structure during the outburst \citep{2017MNRAS.470...48V}.
In this work, we report on two epochs of simultaneous X-ray/IR photometry at high time resolution (i.e. sub-second) for this source.
%This paper is structured as follows: in Section \ref{sec:obs}, we describe the observations analysed, the data reduction and data processing techniques that preceded the analysis; the Section \ref{sec:analysis}, contains data analysis carried out in both the time and frequency domains; we discuss about the results of the first and second epochs within Section \ref{sec:discussion}, underlining the possibility of the presence of a jet in the source; finally in the conclusions, Section \ref{sec:conclusion}, we summarize the results and their interpretation.

\section{Observations}
\label{sec:obs}
%\begin{table}
%	\centering
%	\caption{Infrared and X-rays observations of SWIFT J1753.5--0127}
%	\label{tab:obs}
%	\begin{tabular}{ccc} % four columns, alignment for each
%		\hline
%		& VLT & X-rays \\
%		 Date & Time Interval (UT) & Time Interval (UT)  \\
%		\hline
%     15 Aug 2008 & 02:40 -- 05:35 & 02:50 -- 04:58 \\
%%       & RXTE &15 August 2008 & 02:50 - 04:58\\
%
%%		Epoch 2&&&\\
%       
%     10--11 Sep 2012 & 23:54 -- 02:43 & 17:52 -- 04:24 \\
%%       & XMM-Netwon &10-11 September 2012 & 17:52 - 04:24 \\
%		\hline
%	\end{tabular}
%\end{table}

\begin{table}
	\begin{center}
	\caption{Infrared and X-rays observations of SWIFT J1753.5--0127}
	\label{tab:obs}
	\begin{tabular}{cccc} % four columns, alignment for each
		\hline
		MJD & Date & \multicolumn{2}{c}{Time Interval (UT)} \\
		& & Infrared & X-rays \\
		\hline
    54693 & 15 Aug 2008 & 02:40 -- 05:35 & 02:50 -- 04:58 \\
    56180-1 & 10--11 Sep 2012 & 23:54 -- 02:43 & 17:52 -- 04:24 \\
		\hline
	\end{tabular}
        \end{center}
\end{table}

\subsection{Infrared data}

We observed Swift J1753.5-0127 in the IR band for $\sim 10$~ks from ESO's Very Large Telescope at Paranal Observatory on 15 August 2008 (ESO program 281.D-5034) and during the night between the 10 and 11 September 2012 (ESO program 089.C-0996): we will refer to these dates as the first and second epoch respectively. The data were acquired using the $K_{\rm S}$ filter with the \textit{Infrared Spectrometer And Array Camera} (ISAAC) \citep{1998Msngr..94....7M} mounted on the 8.2-m UT1/Antu telescope. The detector was windowed to $256 \times 256$~pixels to reduce the time resolution to $62.5$~ms. The data acquired by the instrument were stored in cubes, i.e. groups of frames ($N_{\rm frames}=995$) collected consecutively over a given time interval (i.e. $\sim$62~s), with $\sim$3~s-long gaps in between. The weather conditions of both observations were similar, with an average seeing around 0.7\lq\lq. The absolute time accuracy is of the order of 10 milliseconds (the readout time of the detector).

The chosen $38\arcsec\times 38\arcsec$ field of view contains the target, a brighter ‘reference’ star ($K_{\rm S}=13.19 \pm 0.03$), located $28.2\arcsec$ south of our target (which we used to reduce the impact of atmospheric turbulence on our light curves), and two faint 'comparison' stars ($K_{\rm S} = 16.12 \pm 0.07$ and $K_{\rm S} = 16.68 \pm 0.11$, located about $15\arcsec$ south-west and south-east of our target, respectively) which we used to optimize the extraction parameters. %as well as to model the high-frequency noise in the Fourier domain.
The ULTRACAM pipeline\footnote{ \url{http://deneb.astro.warwick.ac.uk/phsaap/software/}.} was used for the data reduction. %The detector window used in the two epochs is exactly the same, but the telescope pointing direction and angle differ slightly so that the four objects fall in different detector positions. This allowed us to use an average image from the 2012 epoch to improve the calibration of the 2008 epoch data, and vice versa. We performed fixed-aperture photometry of all the sources in the field of view (target, reference star and two comparison stars). For each source in the field of view, we used a circular region to extract the counts of the source and an annular region centred around the source, to extract the counts of the background.

We found an average magnitude for our target during the first (second) epoch of $K_{\rm S} = 14.95 \pm 0.05$\,mag ($15.05 \pm 0.06$\,mag), corresponding to a flux of $F = 0.70 \pm 0.03\,(0.64 \pm 0.04)$~mJy. We did not correct these values for interstellar absorption. 

% For the second epoch we estimated an average magnitude of $K_s = 15.0 \pm 0.3 \, mag$, which corresponds to an average flux of $F = (0.7 \pm 0.2) \times 10^{-29} \, W m^{-1}Hz^{-1} = 0.7 \pm 0.2 \, mJy$.

Before performing the timing analysis, the IR light curves were barycentred to the Solar System Barycentre using a custom MATLAB software (Ambrosino et al., in prep.). Given the relevance of this correction and the possible impact on our results of any inaccuracy, we cross-checked our barycentric correction with another software \footnote{\url{https://astroutils.astronomy.osu.edu/time/utc2bjd.html}} and found no significant differences. 

\subsection{X-ray data}

We observed Swift J1753.5--0127 in the X-ray band simultaneously with the IR data in both epochs.
The total exposure of the first observation was $\sim 3.5$~ks, obtained with the Proportional Counter Array (PCA) on board the \textit{Rossi X-ray Timing Explorer} (RXTE) satellite (ObsIds: 93105-01-57-00, 93105-01-57-01). Spectral fitting with a simple power-law model results in a 2–10 keV unabsorbed flux of F$_X\sim10^{-9}$ ergs/s/cm$^2$ (corresponding to a luminosity L$_X\sim7.7\times10^{36}$ ergs/s at 8~kpc).
A light curve was extracted in the $2-15$~keV energy range with $15.625$~ms (1/64 s) time resolution, using standard HEADAS 6.5.1 tools. 

In the second epoch, the source was observed for $36$~ks with the Epic-pn (PN in the following) camera on board the \textit{XMM-Newton} satellite operated in TIMING mode (ObsID: 0691740201). We filtered and screened the PN data using the Science Analysis Software (\textsc{SAS}, \citealt{Gabriel2004}) v. 19.0.0 with the up-to-date calibration files. We searched for possible intervals of flaring particle background by extracting the single event (pattern=0) high-energy (10.0--12.0~keV) light curve, but we found none. We then filtered the PN data by retaining only events with pattern $\le 4$ (single and double pixel pattern only) and falling in the region RAWX range [30:46]. Finally, we barycentred the PN photon arrival times using the \texttt{barycen} tool and adopting DE-405 solar system ephemeris. Spectral fitting with a simple power-law model results in a 2–10 keV unabsorbed flux of F$_X\sim4\times10^{-10}$ ergs/s/cm$^2$ (corresponding to a luminosity L$_X\sim3\times10^{36}$ ergs/s at 8~kpc). A light curve was extracted with a time resolution of $1$~ms, in the $0.5-10$~keV energy range. We checked that extracting the light curve only above 2~keV does not affect the results significantly, thus we decided to retain the full energy selection to optimise the throughput.

Both X-ray light curves were barycentred to the Solar System barycentre, using standard \textsc{HEASoft} tools. The resulting curves were then rebinned to match and align to the simultaneous IR time series. The absolute time accuracy of RXTE and \textit{XMM-Newton} is 2.5 and 48 $\mu$s respectively \citep{2006ApJS..163..401Jahoda,2012A&A...545A.126Martin-Carrillo}.
%We did this with a custom Python code, which aligned the X-ray new bins to the IR bins. This operation introduces an uncertainty, as the boundaries of the original X-ray bins were not perfectly aligned with the IR bins. Therefore, some bins of the X-ray curve fall on the boundary between two consecutive IR bins, which forces the tool to choose to which IR bin to assign the "contended" X-ray bin.
%The time shift introduced between the two signals changes over time, due to the fact that the time difference between consecutive IR data cubes is not constant and is not an integer multiple of the X-ray time resolution. However, this uncertainty is negligible, as the measured delays are much larger than the original X-ray time resolution. The mean and standard deviation of the time shift of the X-ray signal introduced by this effect are respectively $0.38 \, ms$ and $0.39 \, ms$ for the first epoch, and $0.014 \, ms$ and $0.29 \, ms$ for the second epoch.

\begin{figure*}
\centering 
\includegraphics[width=0.98\textwidth]{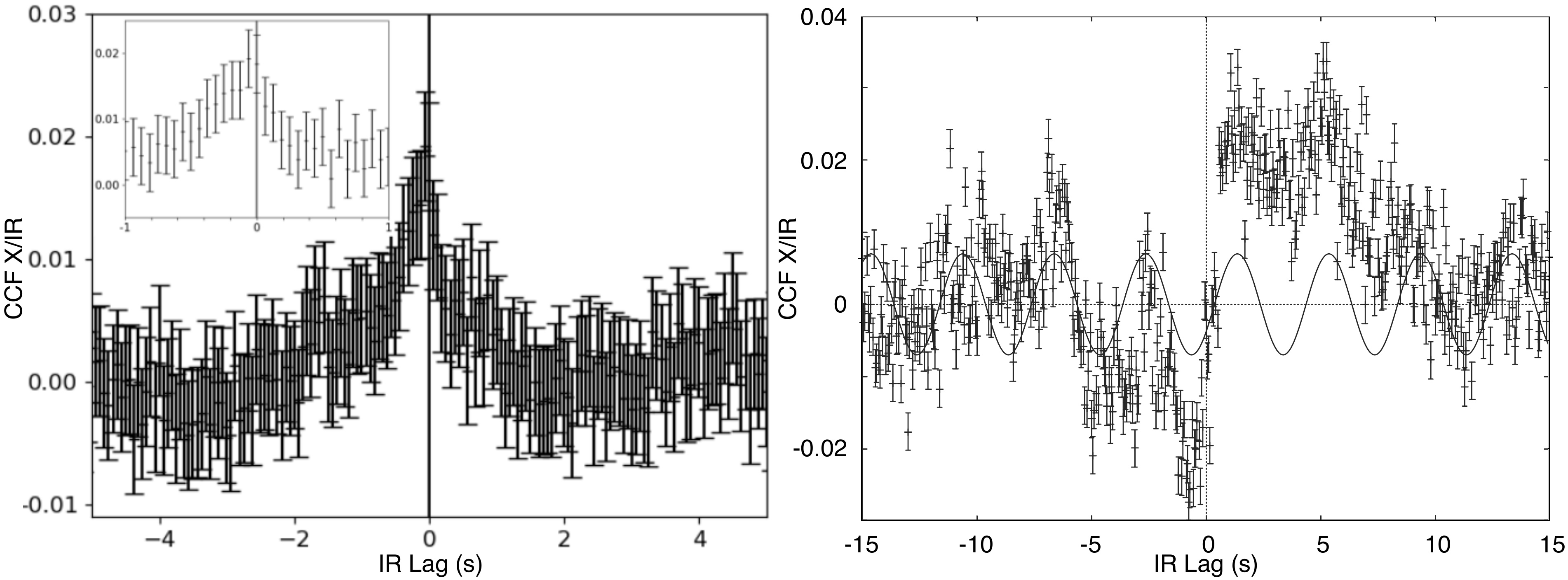}

\caption{The IR/X-ray cross-correlation functions for the first (\textit{left}) and second (\textit{right}) epoch, both calculated and plotted with a 62.5 ms time resolution. Positive lags mean that IR lag X-rays. A clear correlation is detected in the 1st epoch, with a peak at $\approx-0.13$~s. The CCF of the second epoch shows a complex structure, with multiple anti-correlation and correlation features. As shown by the sinusoidal wave (with a period of 4 s), most of the peaks seem to be associated with the QPO at 0.25 Hz visible in the IR PDS (Fig. \ref{fig:fre_dom_1_epo})  }
\label{fig:CCF_X_IR}
\end{figure*}

\begin{figure}
    \centering
    \includegraphics[width=0.4\textwidth]{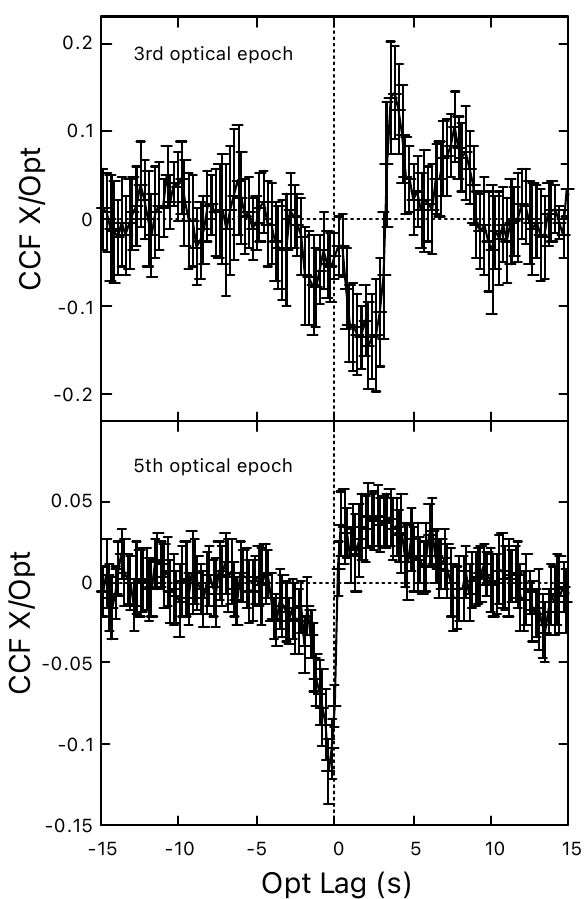}
    \caption{The optical/X-ray cross-correlation functions for the third (\textit{top}) and fifth (\textit{bottom}) optical epoch \cite[from][]{2017MNRAS.470...48V}.}
    \label{fig:CCF_Veledina}
\end{figure}
 
\section{Analysis}
\label{sec:analysis}
\subsection{Cross-correlation function}

To quantify the correlation between the X-ray and IR bands we computed the cross-correlation function (CCF) between the two time series, normalized by the product of standard deviation in each band, using the procedure described in \cite{2010MNRAS.407.2166G}, at the maximum available time resolution of 62.5 ms. The instruments and the barycentre correction method used for this work allow us to measure lags with a timing accuracy of a few tens of ms. The two CCFs (Fig. \ref{fig:CCF_X_IR}) reveal a clear difference between the two epochs. 
The first epoch has a single peak structure that is similar to the ones observed in other sources \citep[e.g.][]{2010MNRAS.404L..21C}, except for the fact that it peaks at slightly negative lags, indicating an infrared lead. We quantify the position of the peak of the lag fitting a single Lorenztian function to the CCF between $-2$~s and 2~s, obtaining a peak lag of -0.13$\pm$0.03~s. The second CCF, instead, has a more complex structure: an anti-correlation dip at negative lags between about $-6$~s and 0~s is followed by a peak of correlation between $\sim 0$~s and $\sim 10$~s. Both the dip and the peak are structured in what appear to be multiple sub-peaks. A more careful look reveals that these sub-peaks are consistent with being equally spaced, with a periodicity of 4~s. We tested this by fitting a sinusoidal function to the CCF, shown in Fig. \ref{fig:CCF_X_IR}. This is consistent with the presence in the two bands of a correlated quasi-periodic oscillation at about 0.25 Hz, which is confirmed by the power spectral analysis of the infrared time series (see next sub-section).
In both epochs, we assessed stationarity by dividing each exposure into two halves. The CCFs calculated independently for the two halves of each epoch do not reveal any significant difference, confirming that the time series are stationary.

\subsection{Power spectral analysis}

To evaluate the Fourier cross-spectral products we followed the procedure described by \citet{2014A&ARv..22...72U}. For both epochs, we computed the discrete Fourier transform using 512 bins per segment and a rebinning logarithmic factor of 1.2 (each bin is 20\% longer than the previous one). 

\begin{figure*}
\centering 
\includegraphics[height=0.47\textwidth,angle=-90]{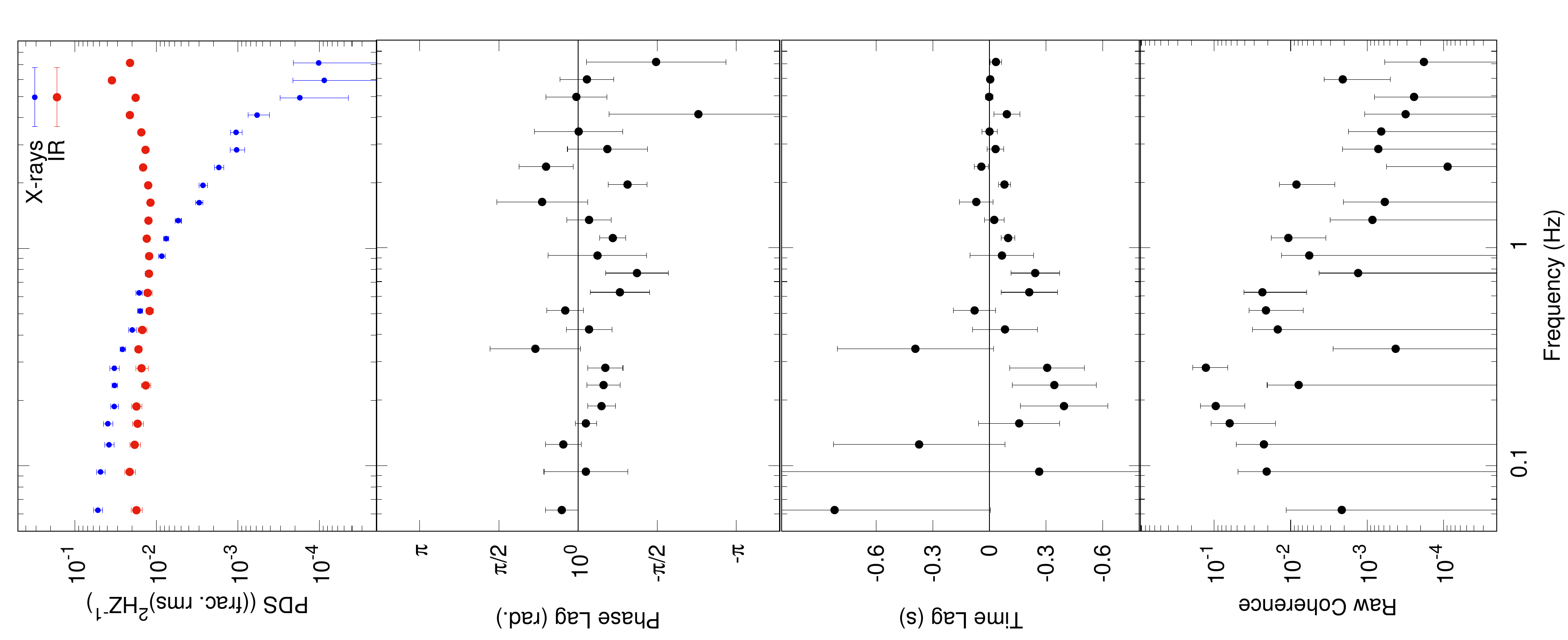}
\includegraphics[height=0.47\textwidth,angle=-90]{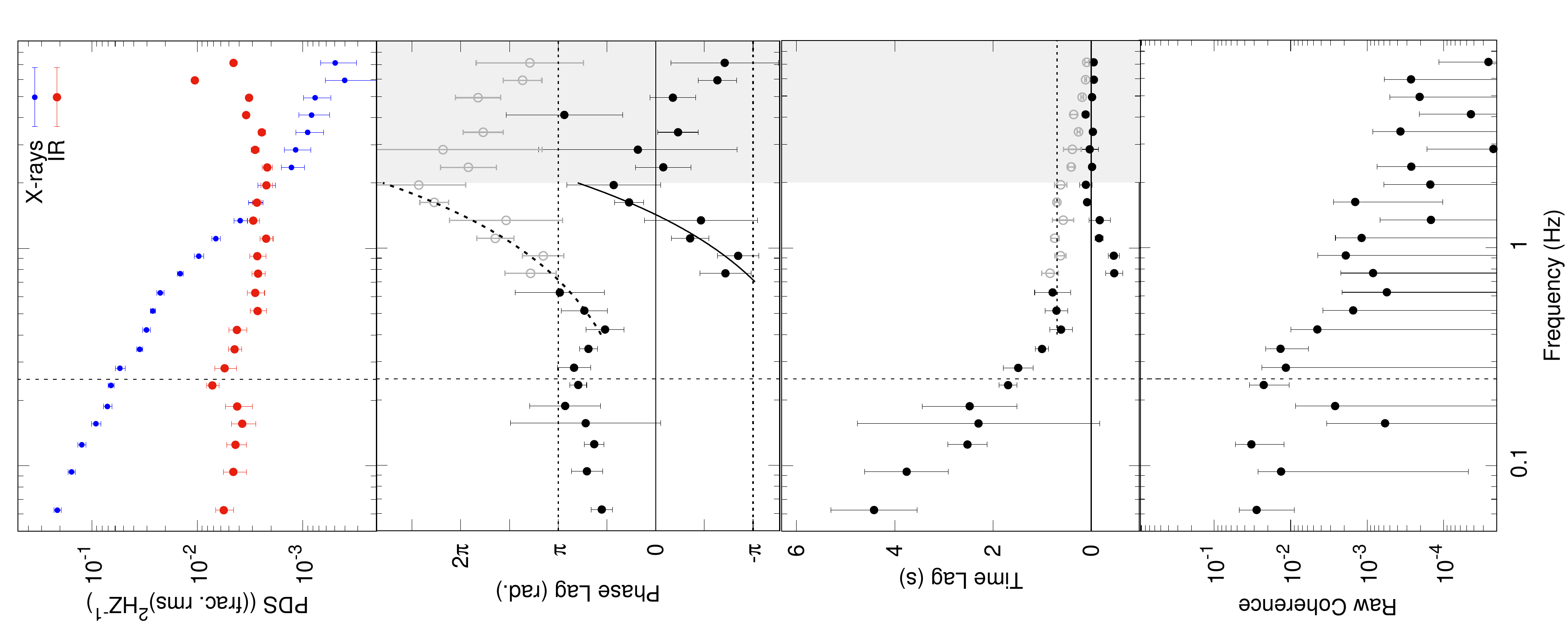}
\caption{Fourier Domain Analysis for the first (\textit{left}) and second (\textit{right}) epoch. \textit{Top Panels:} X-ray (blue) and IR (red) PDS. Counting noise was subtracted from the X-ray PDSs but not from the IR, owing to the instrumental features present in the IR PDS (see Appendix \ref{sec:appendix}). %This is because a high-frequency blue noise component,  visible in the IR PDS. This feature is instrumental as it was also visible in the signal from the comparison stars in the field. 
Evidence for a quasi-periodic oscillation can be seen at  $\approx$0.25 Hz in the PDS of the second epoch, marked by the vertical dotted line.
\textit{Second Panels:} Phase lag spectrum. Positive lags mean that IR lags X-rays. While the first epoch generally has a negative lag, the second is dominated by a strong positive lag component. A clear discontinuity is seen in the second epoch, caused by phase wrapping. The correction for phase wrapping is shown via grey points. The curved lines mark the estimated constant time lag (with and without phase wrapping correction). It is apparent that at $\sim 2$ Hz, further phase wrapping occurs, randomising the lags at higher frequencies (grey area). 
\textit{Third Panels:}  Time lags as a function of Fourier frequency. The horizontal dotted line marks the estimated constant time lag of $\sim 0.7$ seconds between 0.4 and 2 Hz, after correcting for phase wrapping (grey points).
\textit{Bottom Panels:}  Raw coherence as a function of the Fourier frequency. }

\label{fig:fre_dom_1_epo}
\end{figure*}

The X-ray and IR power density spectra (PDSs) in fractional squared root-mean-square units \citep{1991ApJ...383..784Miyamoto} are shown in Fig.~\ref{fig:fre_dom_1_epo}. Counting noise was subtracted from the X-ray PDSs but not from the IR PDSs. {This is because we found in the PDS of both target and comparison star a blue noise component, along with spurious instrumental spikes, at frequencies higher than $\approx$ 1Hz  (see Fig. \ref{fig:plot_pds_comparison_star} and the discussion in the Appendix \ref{sec:appendix}).}  %(present also in the time series of the comparison stars in the field), which makes it difficult to fully quantify the contribution of the different (counting and read-out) noise components (see Fig. \ref{fig:plot_pds_comparison_star} and the discussion in the Appendix \ref{sec:appendix}). %{ \bf Another source of noise is also evident in the 4 - 6 Hz range, appearing as a pair of peaks at $\sim$ 4 and $\sim$ 6 Hz. The presence of this component has already been identified in \citet{2010MNRAS.404L..21C} and \citet{2018MNRAS.477.4524V_vincentelli}, where it is shown to be of instrumental origin (it is also present in the PSD of the comparison star, see Fig. \ref{fig:plot_pds_comparison_star} and Fig. 2 in \citet{2018MNRAS.477.4524V_vincentelli}), and it is negligible in the correlation study as it is present only in the IR signal.} 
Because of this, and also owing to their low statistics, we do not model the PDSs. Nevertheless, we note that an uncorrelated noise component will not affect the measurement of the lags, only the amplitude of the CCF (or equivalently the Fourier cross-spectral coherence). We note however that the IR PDS of the second epoch (Fig.~\ref{fig:fre_dom_1_epo} top right panel) shows an excess at 0.25 Hz. This frequency is consistent with the periodicity observed in the CCF (Fig.~\ref{fig:CCF_X_IR}, right panel). We quantify this feature by fitting it with a Lorentzian and modelling the surrounding continuum with a simple power-law.  We found that the QPO is significant at a  $\approx2.5 \sigma$ level, with a fractional rms of $2.0\pm0.4\%$. We note the possible presence of a modulation in the CCF that would be consistent with being associated with the QPO. We did not perform any deeper statistical analysis of the significance of this feature, as it would go beyond the scope of this work.

\subsection{Cross-spectral analysis}

We computed the cross-spectrum for both epochs, using the recipe reported in \cite{2014A&ARv..22...72U}, keeping the same number of segments and rebinning factor as for the PDS described in the previous section. We then extracted the time-, phase-lags and raw coherence for the two epochs, shown in Fig.~\ref{fig:fre_dom_1_epo}. Positive lags imply that the IR band lags the X-rays. We did not attempt to compute the intrinsic coherence owing to the presence of a blue component in the PDS which prevents a reliable estimate of the overall noise. The strong differences observed in the CCFs of the two epochs are reflected also in the frequency domain.

During the first epoch (Fig. \ref{fig:fre_dom_1_epo}, left panel) the phase lags are negative over nearly the whole frequency range, albeit with large uncertainties that make almost all data points compatible with zero lag. We note however that the centroid and width of the CCF peak are roughly consistent with the IR lead that is observable at $\nu \sim 1$ Hz, as expected. 

During the second epoch (Fig. \ref{fig:fre_dom_1_epo}, right panel), instead,  the phase lag spectrum has a different structure. The lags are nearly constant at $\sim 2$ rad at low frequencies, up to $\sim0.7$~Hz where the lags change sign abruptly. Such a discontinuity suggests that the signal underwent phase wrapping. This is confirmed by shifting by $2\pi$ the phase lags above 0.7~Hz (grey points in the figure), which reveals a smooth evolution up to at least 2 Hz, where probably further phase wrapping appears and randomizes the lags. 

\section{Discussion}
\label{sec:discussion}
We have analysed two epochs of simultaneous IR and X-ray fast photometry of the black-hole transient Swift J1753.5--0127 during the late stages of its very long 2005 discovery outburst. In both epochs we detect correlated variability between the two time series, but with remarkably different properties. During the first epoch, we find that the IR variability leads the X-ray variability by $\sim130$~ms, as evident both in the CCF and in the frequency-dependent lags. In the second epoch, instead, the correlation between the time series is more complex, with different lags appearing at different frequencies, suggesting the presence of multiple components.

Multiple epochs of simultaneous optical and X-ray fast photometry of Swift~J1753--0127 were reported by several authors \citep{2008ApJ...682L..45D_durant,2009MNRAS.399..281H,2009MNRAS.392..309D,2011MNRAS.410.2329D_durant}, and their complex behaviour was summarised and discussed in \citet{2017MNRAS.470...48V} in the context of a self-consistent physical scenario with multiple components and several parameters: an expanding hot accretion flow with multiple synchrotron and Comptonization components, a variable contribution from thermal disc reprocessing and -- whenever present -- a contribution from a QPO.
The five epochs discussed by these authors are shown in Fig.~\ref{fig:light_curve}.
In the following, we discuss possible interpretations of the newly found behaviour and test the hot flow scenario against the new data.

\subsection{First Epoch}

We observe the infrared variability to lead over the X-ray variability, thus we can safely rule out a jet and a reprocessing origin for the IR variable emission in this epoch, as in both cases an infrared lag would be expected.
Our first epoch occurred only five days after the fifth epoch considered in \citet{2017MNRAS.470...48V}. The position of the source in the hardness-intensity diagram in the two epochs did not change significantly between the two epochs (see Fig. \ref{fig:light_curve}, left panel), thus we can safely assume the physical and geometrical conditions did not change substantially. However, their CCF is very different from ours (see Fig.~\ref{fig:CCF_Veledina}, bottom panel): they measure an anti-correlation at negative lags and a correlation at positive lags, while we observe a correlation at negative lags. \citet{2017MNRAS.470...48V} interpret the correlation at positive lags in terms of disc reprocessing, while the anti-correlation at negative lags is interpreted in terms of a synchrotron (optical) self-Compton (X-ray) scenario from a magnetised corona. In this scenario, the anti-correlation between synchrotron and self-Compton emission originates in fluctuations in the magnetic field causing a pivoting in the spectrum around the energy that divides the two physical processes \citep{2011ApJ...737L..17Veledina}. We do not observe any sign of disc reprocessing in the infrared, which is perhaps not surprising given the different wavelengths (the infrared-emitting region of the disc is expected to be much further away than the optical-emitting region). We do not observe any anti-correlation either, predicted by the magnetised corona scenario because of the pivoting of the spectrum. However, in that scenario, another pivoting is predicted \citep[see Fig. 1 in][]{2009ApJ...690L..97P}: below the self-absorption break, the synchrotron emission from the corona is expected to correlate with the self-Compton emission. Thus, a positive correlation between IR and X-ray emission could be expected as long as the self-absorption break is at a shorter wavelength than IR (thus $\lambda \lesssim 2 \mu$m, as we performed our observations with the $K_{\rm S}$ filter). This is in principle feasible, although it requires a surprising fine-tuning: the break must also be at a longer wavelength than optical (thus $\lambda \gtrsim 0.7 \mu$m, as the observation in the fifth optical epoch was performed with the {\it r} filter), to explain the anti-correlation found in \citet{2017MNRAS.470...48V} only five days earlier. 
This leaves a very narrow wavelength range. Breaks are often observed in the spectral energy distributions in the OIR wavelength range, and existing models for the magnetised inflow do predict a break at least around those wavelengths. On the other hand, the OIR spectral energy distribution of Swift J1753.5-0127 (or at least its non-variable component) has been well fitted by a (non-irradiated) disc thermal spectrum, with evidence for an IR excess component that takes over at longer wavelengths than H-band \citep{2014ApJ...780...48Froning,2014ApJ...788..184Wang}. Given the lack of a perfect simultaneity between our observation and those of \citet{2017MNRAS.470...48V}, it is difficult to draw more definite conclusions.

\subsection{Second Epoch}
The CCF in our second epoch has a complex shape, which is more difficult to interpret. We observe multiple anti-correlations at negative lags and an equally structured correlation at positive lags. The CCF is somewhat similar to the CCF observed in the third and fifth optical epochs by \citet{2017MNRAS.470...48V} (see Fig. \ref{fig:CCF_Veledina}), about seven and four years earlier respectively. Both CCFs were modelled by \citet{2017MNRAS.470...48V} in the context of the expanding hot flow model described above. Thus, it is natural to expect that the same model might succeed in describing our CCF. However, some of the complex structures we observe are most probably caused by the presence of a QPO (Fig. \ref{fig:CCF_X_IR}), which was not present in the fifth optical epoch. A QPO at a very similar frequency was instead present in the third epoch and apparent in the CCF, which looks surprisingly similar to the one we measured in our second epoch, although it was shifted by 2 seconds toward positive (optical) lags. As a result of these and further differences (including the X-ray brightness), the two epochs were modelled with two rather different sets of parameters. The phase lags of those two epochs are completely different from each other, demonstrating that the CCFs alone are not always sufficiently informative. A QPO at lower frequencies was present in the fourth optical epoch, and apparent in the CCF \citep{2015MNRAS.454.2855Veledina}.

Given the limitation of the technique and the complexity of our CCF, cross-spectral analysis here can help us separate different components at different timescales, revealing the possible role of different processes. Looking at Fig.~\ref{fig:fre_dom_1_epo} we can identify two different behaviours/regimes in the lags: at low frequencies, the phase lags are consistent with being constant at around $\sim 2$ radians, corresponding to a lag of a few seconds decreasing with frequency. The amplitude of these lags and the frequency intervals in which they appear suggest they could be consistent with the model proposed by  \citet{2017MNRAS.470...48V}. At high frequencies instead, above $\sim 0.4$~Hz, the phase lags increase with frequency, corresponding to a constant lag at $\sim 0.7$~s. By integrating the phase lag between 0.4-0.9 Hz (i.e. where the phase-wrapping effect is still not dominant) we obtain an IR lag of 0.72$\pm$0.16s. A constant lag at these frequencies, albeit at a shorter value of the order of 0.1~s, has been observed in other BHTs and associated with the jet (see the following Subsection). For frequencies above $\sim 2$~Hz, there is evidence for intense phase wrapping that makes it impossible to extract useful information. No clear feature can be seen in the lags at the frequency of the QPO ($\simeq0.25$~Hz), owing to the low significance of the QPO in the PDSs. For the same reason, we cannot quantify the QPO lag from the modulation apparent in the CCF, considering also the complexity of the CCF itself. We note however that qualitatively the modulation in the CCF suggests that the lag is small, perhaps consistent with the zero QPO lag measured by \citet{2015MNRAS.454.2855Veledina} in the optical band.

\subsection{A jet in a radio-quiet source?}

The most evident property of the correlated variability in Swift J1753.5--0127 is its complex evolution. Only two of the five epochs of X-ray/optical variability reported by \citet{2008ApJ...682L..45D_durant,2009MNRAS.399..281H,2009MNRAS.392..309D,2011MNRAS.410.2329D_durant} and discussed by \citet{2017MNRAS.470...48V} are similar to each other, while the others show remarkably different features. This complexity is confirmed by our results. Not only do our two epochs of X-ray/IR variability differ enormously from each other, but they are also rather different from all the optical epochs, even when they are very close in time. Even when the CCFs show some similarities, the time lags reveal important differences.

\begin{figure}
\centering 
\includegraphics[width=0.5\textwidth]{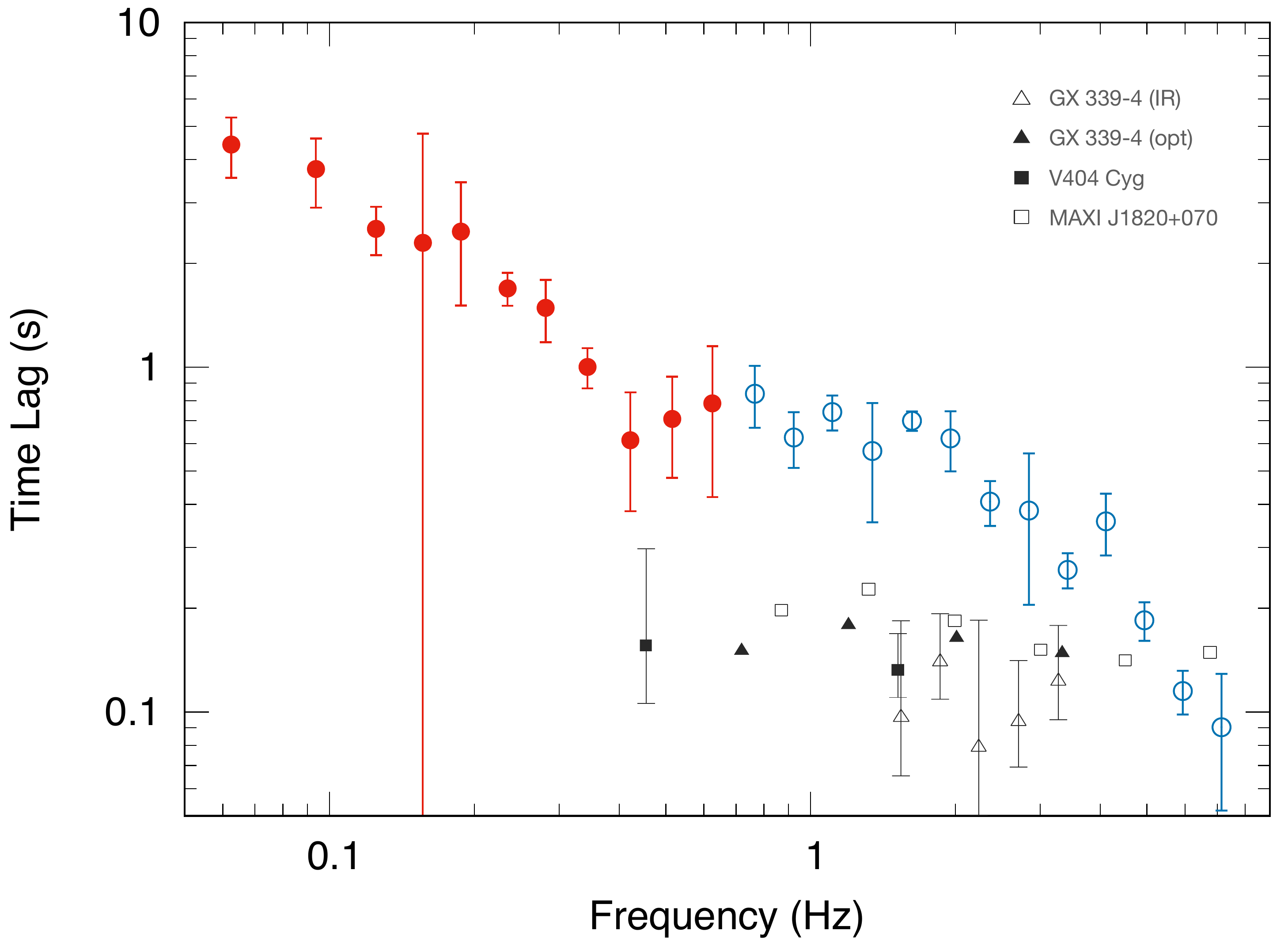}
\caption{X-ray/IR time lags for Epoch 2. The red-filled points are the original data points, while the empty blue circles are the time lags computed assuming a 2$\pi$ shift, i.e. after taking into account the phase wrapping effect. Two components are evident, with a transition at around 0.4 Hz. In dark grey, we report previous measurements of a constant time lag at frequencies around 1 Hz (see text).} 
\label{fig:tlags_2epo}
\end{figure}

While in our first epoch, there is no evidence of any jet contribution to the infrared emission, our second epoch shows time lags somewhat similar to those observed in other BHTs, where they were interpreted as a jet signature. If we compare the time lags from our second epoch (Fig.~\ref{fig:tlags_2epo}) with those reported for GX~339--4 in \citet[][Fig. 18]{2010MNRAS.407.2166G} and \citet[][Fig. 2]{2019ApJ...887L..19V_vincentelli}, for V404~Cyg in \citet[][Fig. S2]{2017NatAs...1..859G_gandhi} and for MAXI~J1820+070 in \citet[][Fig. 2]{2019MNRAS.490L..62P_paice}, the similarities are striking. In all cases, a clear component with a constant lag at $\sim 0.1-0.2$ seconds can be identified at frequencies around 1 Hz. 
%In all cases, this component extends up to the highest measurable frequencies, either limited by the time resolution of the dataset or by phase wrapping, suggesting that the very same component might extend up to higher frequencies.
This component has been associated, very securely in some cases, by analogy in others, with a compact jet. Thus, it is natural to consider the possibility that we have detected variable jet emission also in Swift J1753.5--0127. If this is the case, the questions remain of why is the measured lag 0.7~s instead of the ``usual'' 0.1--0.2~s, and why no jet contribution is detected in our first epoch, despite the very similar X-ray hardness. We note that the X-ray luminosity during our second epoch was very similar to (a factor of $\sim2$ brighter than) the X-ray luminosity at which the 0.1-s lag was measured in GX~339-4 \citep{2010MNRAS.404L..21C}. 

\begin{figure}
\centering 
\includegraphics[width=0.5\textwidth]{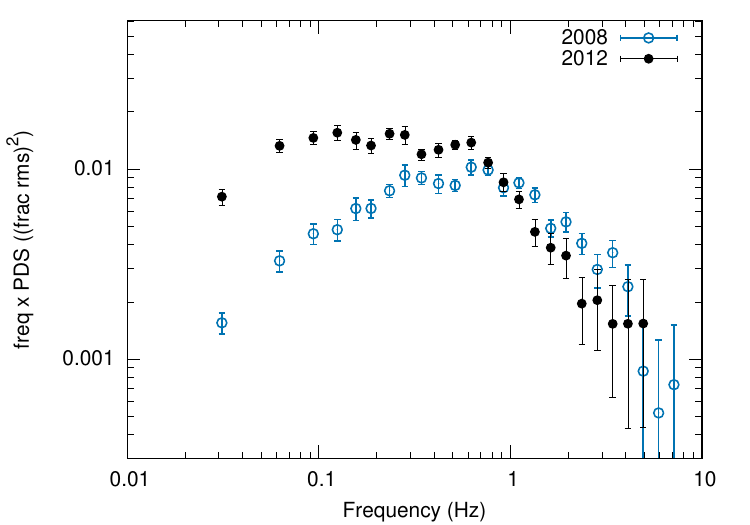}
\caption{Comparison between the X-ray PDSs measured during our 2008 and 2012 observations. A clear change of the slope is observed below the high-frequency break.} 
\label{fig:pds_x_comp}
\end{figure}

The longer jet lag in Swift J1753.5--0127 might be related to the well-known peculiar properties of the jet in this source, which is one of the so-called radio-quiet black-hole transients \citep{2011MNRAS.413.2269S,2012MNRAS.423..590G,2018MNRAS.473.4122E,2018MNRAS.478.5159M}. This definition comes from the behaviour of this source in the radio/X-ray plane, where it lies below the ``standard'' track \citep{2010MNRAS.406.1471S}, and suggests that the jets in these sources have different radiative properties, and are perhaps weaker than in ``standard'' sources. Thus, it would not be too surprising if the jet in Swift J1753.5--0127 also had different variability properties. This could be somewhat confirmed by the fact that, in at least one case, the jet break in Swift J1753.5-0127 has been constrained to be at frequencies lower than $3.6\times10^{12}$ Hz \citep{2015ApJ...808...85Tomsick}, at least an order of magnitude lower than in GX 339-4 \citep{2011ApJ...740L..13Gandhi} and in several other BHTs \citep{2013MNRAS.429..815Russell}. 
The differences between the two epochs could be due to two different reasons: on the one hand, the second epoch has a lower count rate than the first epoch. The jet could have appeared as the source was heading towards quiescence, while it was not present years earlier in a brighter state. Several radio-quiet black-hole transients have shown a transition back from the radio-quiet branch to the standard branch in the radio/X-ray plane when heading towards quiescence \citep[e.g.][]{2011MNRAS.414..677C}, suggesting the jet becomes stronger at low accretion rates. Evidence for this happening also in Swift J1753.5--0127 was reported \citep{2016AN....337..485K}.
On the other hand, the X-ray variability in the two epochs differs substantially (Fig.~\ref{fig:pds_x_comp}), with a clear change in the slope of the power spectrum below the high-frequency break. The two shapes are consistent with the typical PDS observed in the hard states \citep{2005A&A...440..207B}. A link between the shape of the X-ray PDS and the radio properties have been recently suggested for GRS 1915+105 \citep{mendez2022} and GX 339-4 \citep{zhang_yuexin2024}.   In the context of the jet internal-shock model \citep{2018MNRAS.480.2054M_malzac}, it has been shown that such a difference in the slope of the PDS would correspond to different jet spectral properties in the two epochs, with the self-absorption break shifting by as much as four orders of magnitude in wavelength \citep{2014MNRAS.443..299M_2014}. Thus, again it would not be too surprising if, in the second epoch, the jet is much brighter in the infrared than in the first epoch. This could in principle also be related to a variable jet speed with luminosity, as it has been suggested by \cite{2015MNRAS.450.1745Russell} to explain the peculiar behaviour in the radio/X-ray plane of MAXI J1836-194 (a system though with most probably a lower inclination than Swift J1753.5-0127). As the IR flux was similar in the two epochs, a different jet brightness in the IR would imply also a variable contribution from a different component.

\section{Conclusions}
\label{sec:conclusion}
We have observed a complex behaviour in the correlated IR/X-ray fast variability of the BHT Swift J1753.5-0127 during its 10-year-long 2005 discovery outburst. In the first of our two epochs, the data can be interpreted in terms of synchrotron-self-Compton emission from a magnetised hot flow. In the second epoch, the data reveal a more complex context. The low-frequency behaviour is consistent with a combination of disc reprocessing and a magnetised hot flow. However, the constant lags at $\sim$0.7 seconds at high frequencies are reminiscent of the constant lags observed in a similar frequency range at $\sim$0.1 seconds in other sources. These constant lags are usually considered a signature of O-IR synchrotron emission from a compact jet lagging by 0.1 seconds the X-ray emission from the inflow, which suggests a similar interpretation for SWIFT J1753.5-0127. The longer lag we measure could be due to the different radiative properties of the jet in this source.  The complexity of the behaviour, the lack of broader multi-wavelength data and the overall paucity of datasets make these interpretations only tentative. These results underline the need for denser campaigns of strictly simultaneous multi-wavelength fast photometry to reach a broader and deeper understanding of the complex variable spectral properties of black-hole transients.

\section*{Acknowledgements}

Based on observations collected at the European Southern Observatory under ESO programmes 281.D-5034 and 089.C-0996.

PC, FMV and all the Authors acknowledge the long-term contribution of Tomaso Belloni to this project. Tomaso sadly passed away in August 2023 and will be sorely missed.

The Authors thank the Team Meeting at the International Space Science Institute (Bern) for fruitful discussions and were supported by the ISSI International Team project \#440.
FMV acknowledges support from the grant FJC2020-043334-I financed by MCIN/AEI/10.13039/501100011033 and Next Generation EU/PRTR, as well as from the grant PID2020-114822GB-I00. PC acknowledges financial support from the Italian Space Agency and National Institute for Astrophysics, ASI/INAF, under agreement ASI-INAF n.2017-14-H.0. AV acknowledges support from the Research Council of Finland grant 355672. Nordita is supported in part by NordForsk. MI is supported by the AASS Ph.D. joint research programme between the University of Rome "Sapienza" and the University of Rome "Tor Vergata", with the collaboration of the National Institute of Astrophysics (INAF).

%%%%%%%%%%%%%%%%%%%%%%%%%%%%%%%%%%%%%%%%%%%%%%%%%%
\section*{Data Availability}

The ISAAC data are available from the ESO Data Archive (\hyperlink{https://archive.eso.org/eso/eso_archive_main.html}{https://archive.eso.org/eso/eso\_archive\_main.html}). The RXTE data are available from the HEASARC Data Archive (\hyperlink{https://heasarc.gsfc.nasa.gov/docs/archive.html}{https://heasarc.gsfc.nasa.gov/docs/archive.html}). The \textit{XMM-Newton} data are available from the \textit{XMM-Newton} Science Archive (\hyperlink{https://nxsa.esac.esa.int/nxsa-web/}{https://nxsa.esac.esa.int/nxsa-web/}). The MAXI data used in Fig.1 are available from the MAXI website (\hyperlink{http://maxi.riken.jp/top/index.html}{http://maxi.riken.jp/top/index.html}). The BAT data used to estimate the hardness of the source during the second epoch are available from the Neil Gehrels Swift Observatory archive (\hyperlink{https://swift.gsfc.nasa.gov/results/transients/}{https://swift.gsfc.nasa.gov/results/transients/}).

\bibliographystyle{aa} 
\bibliography{biblio}

\listofobjects 

\appendix

\section{Comparison star's power spectrum}
\label{sec:appendix}

\begin{figure}
\centering 
\includegraphics[width=0.5\textwidth]{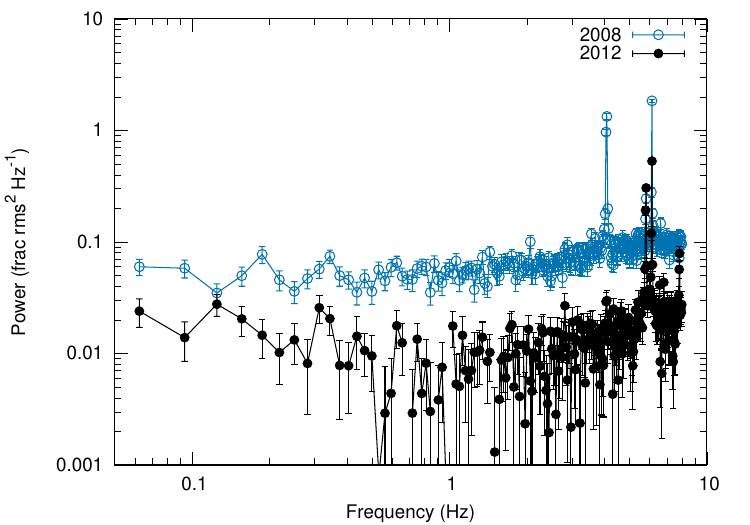}
\caption{Infrared power spectral density of the brighter comparison star during the two epochs of observation.} 
\label{fig:plot_pds_comparison_star}
\end{figure}

To understand the nature of the blue noise present in the IR power spectrum of our target, we also checked the power spectrum obtained from the comparison star. The results for both epochs are shown in Fig. \ref{fig:plot_pds_comparison_star}. It is clear that in both cases there is a source of blue noise which starts dominating above a few Hz. We also notice the presence of spurious peaks at around 4 and 6 Hz (somewhat visible also in the target PDSs in Fig.\ref{fig:fre_dom_1_epo}, top panels). Similar features had already been detected in ISAAC data \citep{2010MNRAS.404L..21C,2018MNRAS.477.4524V_vincentelli}. These peaks are clearly instrumental, while the blue noise is most probably caused by readout noise. No additional feature is observed at lower frequencies, indicating that power measurements below $\sim$ 1 Hz are safe.
%This suggests that the origin of this component is instrumental, most probably caused by readout noise.

\end{document}